\documentclass[a4paper, 9pt, twocolumn]{article}
\usepackage{geometry}                
\usepackage{graphicx}
\usepackage{amssymb}
\usepackage{amsthm}
\usepackage{epstopdf}
\usepackage{mathtools}
\usepackage[shortlabels]{enumitem}
\usepackage{dsfont}
\usepackage{cuted}
\usepackage{sidecap}
\usepackage[usenames,dvipsnames]{color}
\usepackage[thinspace,mediumqspace,Grey,squaren]{SIunits}
\usepackage{bbm}
\usepackage{authblk}
\usepackage[font=small,labelfont=bf ,textfont={small, sf}]{caption}
\usepackage{titlesec}
\usepackage{url}
\usepackage{hyperref}
\titleformat{\subsection}[runin]{\normalfont\bfseries}{\thesubsection.}{3pt}{}

\titleformat*{\section}{\large\bfseries}

\newcommand{\argmin}[1]{\underset{#1}{\mathrm{argmin}}~}

\def\NumReactions{L}
\def\NumSpecies{K}

\newcommand{\Expect}[1]{ \mathbb{ E } \left[ #1 \right]}

\begin{document}

\title{
A Molecular Implementation of the Least Mean Squares Estimator
}

\author[1]{Christoph Zechner\thanks{czechner@ethz.ch}}
\author[2]{Mustafa Khammash\thanks{mustafa.khammash@bsse.ethz.ch}}
\affil[1]{Department of Biosystems Science and Engineering, ETH Zurich, Basel, Switzerland}
\renewcommand\Authands{ and }

\date{}

\maketitle

\newpage{}


{\noindent \sf \fontsize{9}{0} \selectfont \textbf{
Abstract -- In order to function reliably, synthetic molecular circuits require mechanisms that allow them to adapt to environmental disturbances. Least mean squares (LMS) schemes, such as commonly encountered in signal processing and control, provide a powerful means to accomplish that goal. In this paper we show how the traditional LMS algorithm can be implemented at the molecular level using only a few elementary biomolecular reactions. We demonstrate our approach using several simulation studies and discuss its relevance to synthetic biology.
}}


\section{Introduction}  

Engineered circuits in living cells often exhibit poor robustness and substantial variations from one cell to the next \cite{Gardner2000, Elowitz2000}. In extreme cases, they are found functional in only a small fraction of cells in an isogenic population, while others act unpredictably. A major cause for such behavior is that the biochemical components that constitute a circuit depend on factors in their molecular environment (or \textit{context}) of the cell \cite{Cardinale2012}. For instance, the rate at which a protein is expressed depends on the gene dosage or the number of available ribosomes and so on. 

Recently, progress has been made in taking into account such environmental factors into the modeling and design of molecular circuits \cite{Batt2016, Zechner2014b, Toni2013, Hasenauer2011}. This can tremendously improve the faithfulness of computational models and in turn the predictability of rationally designed circuits. However, in  practical scenarios, the origins and properties of potential disturbances are barely known and hard to anticipate during design time. From this point of view, it seems barely realistic to tune a circuit \textit{in silico} such that it acts robustly under all possible perturbations that it may encounter in the real environment of a cell.

A viable alternative is to employ adaptive design principles in which a circuit continuously senses and adjusts itself to changing environmental conditions. This requires molecular circuits that learn and make inference about their surroundings. A few attempts have been made recently to devise such circuits in the form of chemical reaction networks, for example to perform neural network computations \cite{Qian2011b}, to realize message passing inference \cite{Adams2013} or supervised learning \cite{Lakin2016}.
Along these lines, we have recently proposed a molecular implementation of an optimal filter that allows one to estimate dynamically changing noise signals \cite{Zechner2016}. This estimator was derived under a Bayesian optimality criterion by employing a Kushner-Stratonovich differential equation \cite{Kushner1964}. 

In the present work, we consider another powerful class of estimation schemes that are frequently used in adaptive signal processing and control theory. These schemes -- termed least means squares (LMS) estimators \cite{Haykin1996, Kay} -- iteratively compute the solution of a general least squares problem through a gradient-based parameter search. This iterative structure allows a circuit to estimate unknown quantities in an adaptive fashion by processing measurements in realtime. In this paper we demonstrate how LMS-type estimators can be realized using elementary biomolecular reactions. Our work is related to \cite{Lakin2016}, where the authors have proposed a DNA-based gradient-descent scheme, which is able to learn \textit{static} linear functions. In this work, however, we focus on \textit{dynamical} and possibly stochastic biochemical systems that shall be identified by a molecular LMS estimator. 

\vspace{0.3cm}
The remainder of the paper is structured as follows. In Section \ref{sec:StochChemKin} we introduce the mathematical notation and models required to describe molecular circuits. In Section \ref{sec:LMS} we introduce the concept of LMS estimation and present possible molecular realizations. In Section \ref{sec:CaseStudies} we test the performance of the proposed circuits using several simulation studies and discuss how they may be used in practical applications.




\section{Biochemical Reaction Networks}
\label{sec:StochChemKin}


We consider well-mixed molecular reaction networks comprising $\NumSpecies$ molecular species $\mathbf{Z}=(\mathbf{Z}_{1}, \ldots, \mathbf{Z}_{\NumSpecies})^{T}$ that interact with each other through $\NumReactions$ reaction channels of the form 
\begin{equation}
	\mathbf{Z} \xrightharpoonup{h_{i}(Z(t))} \mathbf{Z} + \nu_{i},
\end{equation}
with $i$ as the reaction index, $Z(t)$ as the abundance of $\mathbf{Z}$ at time $t$, $h_{i}$ as a rate function determined by the law of mass-action and $\nu_{i}$ as the stoichiometric change associated with reaction $i$. Throughout this paper, we follow the convention to denote molecular species as boldface symbols. Note that we use the same symbol also to refer to the circuit that those species constitute.

We describe the time-evolution of $\mathbf{Z}$ as a continuous-time Markov chain (CTMC) that can take into account the inherent randomness of biochemical reactions \cite{VanKampen1992}. It can be shown \cite{Kurtz2011} that the molecular abundance $Z(t)$ satisfies a stochastic integral equation of the form
\begin{equation}
	Z(t) = Z_{0} + \sum_{i=1}^{\NumReactions}  \mathcal{P}_{i}\left(\int_{0}^{t} h_{i}(Z(s))\mathrm{d}s\right) \nu_{i},
	\label{eq:RTC}
\end{equation}
where $ \mathcal{P}_{i}$ is an independent unit Poisson processes describing the firings of reaction $i$. Eq. (\ref{eq:RTC}) is commonly known as the \textit{random time change model}.

Assuming the chemical species to be highly abundant, molecular fluctuations become negligible and eq. (\ref{eq:RTC}) can be approximated by a deterministic rate equation of the form
\begin{equation}
	\frac{\mathrm{d}}{\mathrm{d}t} \tilde{Z}(t) = \sum_{i=1}^{\NumReactions} h_{i}(\tilde{Z}(t)) \nu_{i}
	\label{eq:RRE}
\end{equation}
with $\tilde{Z}(t) \approx Z(t)$. We will make use of equations (\ref{eq:RTC}) and (\ref{eq:RRE}) at a later in this manuscript to model stochastic and deterministic reaction networks, respectively.

\section{A Continuous-Time LMS Algorithm}
\label{sec:LMS}

Suppose a circuit requires knowledge about certain environmental factors $\theta$. For example, $\theta$ could be the number of phosphotases available to the circuit. However, these factors are typically not accessible directly by the circuit but only indirectly through available intermediates $\mathbf{Y}$. In the example above, $\mathbf{Y}$ could be a protein that is targeted by that phosphatase, for instance.

The idea is now to use a second molecular circuit $\mathbf{X}$ that is able to identify the dynamics of $\mathbf{Y}$ through a suitable adaptation scheme. We assume here that $\mathbf{X}$ and $\mathbf{Y}$ are equivalent in their structure but have distinct parameters $\hat{\theta}$ and $\theta$, respectively. The goal of the adaptation scheme is to adjust the parameters $\hat{\theta}$ such as to minimize the discrepancy between the measured output $Y(t)$ and the output of $\mathbf{X}$ (termed $X(t)$). The resulting optimal parameters $\hat{\theta}^{*}$ then represent an estimate of $\theta$. 


A suitable and analytically convenient metric to assess the discrepancy between $Y(t)$ and $X(t)$ is the mean squared error
\begin{equation}
 J(\hat{\theta}) = \Expect{(X(t) - Y(t))^{2}}.
\end{equation}
According to this measure, we seek for the set of parameters that minimizes $J(\hat{\theta})$, i.e.,
\begin{equation}
	\hat{\theta}^{*} = \argmin{\hat{\theta}} J(\hat{\theta}), \label{eq:LMS}
\end{equation}
in which case $\hat{\theta}^{*}$ is referred to as the least mean squared (LMS) estimator. The closed-form solution of this optimization problem can be found in certain specific scenarios, for instance in the case of linear system dynamics \cite{Kay}. In most scenarios, however, (\ref{eq:LMS}) is analytically intractable and one has to minimize $J(\hat{\theta}^{*})$ numerically. Note that iterative schemes may be beneficial even when (\ref{eq:LMS}) is analytically tractable, because it gives $\mathbf{X}$ the flexibility to readapt to changes in $\theta$ as will be shown later in this manuscript. 

A common strategy to minimize $ J(\hat{\theta})$ is to employ a gradient-based method that, at each iteration, moves the parameters $\hat{\theta}$ along the direction of the steepest descent, giving rise to the well-known LMS algorithm. This algorithm is usually used in a discrete-time scenario, for instance when operated on a digital signal processing unit. In such case, at each time iteration $n$, the algorithm would update the parameters using the relation
\begin{equation}
\hat{\theta}^{n+1} = \hat{\theta}^{n} - \alpha(n) \frac{\partial}{\partial \hat{\theta}}  J(\hat{\theta}) \Bigg|_{\hat{\theta}^{n}}
\end{equation}
with $\alpha(n)$ as a tuneable step-size. The choice of the latter usually involves a tradeoff between the rate of convergence and the steady sate error of the scheme (i.e., \textit{excess error}). 

Since we consider continuous-time dynamical systems, we seek for an infinitesimal variant of the LMS algorithm \cite{Karni1989}, i.e., 
\begin{equation}
\frac{\mathrm{d}}{\mathrm{d}t}\hat{\theta}(t) = - \alpha(t) \frac{\partial}{\partial \hat{\theta}}  J(\hat{\theta})  \Bigg|_{\hat{\theta}(t)},
\label{eq:ContLMS}
\end{equation}
in which case $\alpha(t)$ can be understood as a rate at which the scheme adapts.

\begin{figure}[ht!]
\begin{center}
	\includegraphics[width=0.65\columnwidth]{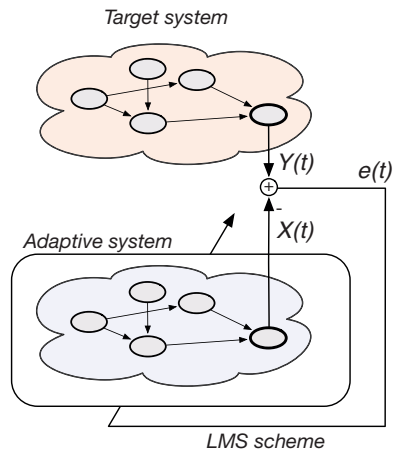}
\caption{Schematic illustration of an adaptive molecular circuit. The goal is to construct a biomolecular circuit $\mathbf{X}$ that adjusts its parameters to mimick those of a target system $\mathbf{Y}$. While those parameters are inaccessible by the circuit, it can measure the output $Y(t)$ of the target system. This output is compared to the output of $\mathbf{X}$ to construct an error signal $e(t)$, which is in turn used to find the optimal parameters of $\mathbf{X}$.}
\label{fig:LMSScheme}
\end{center}
\end{figure}

For the sake of simplicity, we make a few simplifications related to $\mathbf{X}$ and $\mathbf{Y}$ before deriving the scheme. A more general scenario, however, will be subject of a future manuscript. First, we assume that $\theta$ (and correspondingly $\hat{\theta}$) contains only a single parameter that needs to be identified from $Y(t)$. Furthermore, we allow only $Y(t)$ to be corrupted by molecular noise, and reactions associated with $X(t)$ are assumed to evolve deterministically (e.g., through appropriate rescaling of the associated reaction rates).  

Under these assumptions, the gradient of $J(\hat{\theta})$ is given by
\begin{equation}
\begin{split}
\frac{\partial}{\partial \hat{\theta}}  J(\hat{\theta})  &= \frac{\partial}{\partial \hat{\theta}}  \Expect{(X(t) - Y(t))^{2}} \\
&= \Expect{\frac{\partial}{\partial \hat{\theta}} (X(t) - Y(t))^{2}} \\
&= \Expect{2(X(t) - Y(t)) \frac{\partial}{\partial \hat{\theta}}  X(t)} \\
&= 2\Expect{(X(t) - Y(t)) \frac{\partial}{\partial \hat{\theta}}  X(t)}.
\end{split}
\label{eq:ContLMS2}
\end{equation}
Since we assume that $X(t)$ is deterministic, (\ref{eq:ContLMS2}) further simplifies to
\begin{equation}
\begin{split}
\frac{\partial}{\partial \hat{\theta}}  J(\hat{\theta})  &= 2\Expect{(X(t) - Y(t)) \frac{\partial}{\partial \hat{\theta}}  X(t)} \\
&= 2 X(t) \frac{\partial}{\partial \hat{\theta}}  X(t) - 2 \Expect{Y(t)} \frac{\partial}{\partial \hat{\theta}}  X(t) \\
&= 2 X(t) S(t) - 2 \Expect{Y(t)} S(t),
\end{split}
\label{eq:ContLMS3}
\end{equation}
with $S(t):=\frac{\partial}{\partial \hat{\theta}}X(t)$ as the sensitivity of $X(t)$ with respect to $\hat{\theta}$. Note that the particular form of this sensitivity depends on $\mathbf{X}$ and how $\hat{\theta}$ enters its dynamics. Specific examples will be given later in Section \ref{sec:CaseStudies}.

Now, plugging eq. (\ref{eq:ContLMS3}) into (\ref{eq:ContLMS}) yields a dynamic equation for the LMS estimator $\hat{\theta}(t)$, i.e.,
\begin{equation}
	\begin{split}
	\frac{\mathrm{d}}{\mathrm{d}t}\hat{\theta}(t) &= - 2 \alpha(t) X(t) S(t) + 2 \alpha(t) \Expect{Y(t)} S(t) \\
	&= - \tilde{\alpha}(t) X(t) S(t) + \tilde{\alpha}(t) \Expect{Y(t)} S(t),
	\end{split}
	\label{eq:ContLMS4}
\end{equation}
with $\tilde{\alpha}(t)=2\alpha(t)$.

\subsection{Online Adaptation.}
Eq. (\ref{eq:ContLMS4}) provides the desired continuous-time solution of the LMS problem. However, in its current form it is not adaptive, meaning that it assumes known (and fixed) statistics of $\mathbf{Y}$ (i.e., the output mean $\Expect{Y(t)}$). In practice, however, such statistics are often unknown and they might also vary over time.
Using LMS estimation, this problem can be bypassed by estimating $\Expect{Y(t)}$ \textit{online} from available measurements $Y(t)$. This way, the required statistics are extracted directly from data, which in turn allows the scheme to readapt when $\theta$ changes. A common and simple approach is to approximate $\Expect{Y(t)}$ by the current value of $Y(t)$ such that
\begin{equation}
	\frac{\mathrm{d}}{\mathrm{d}t}\hat{\theta}(t) = - \tilde{\alpha}(t) X(t) S(t) + \tilde{\alpha}(t) Y(t) S(t)
	\label{eq:LMSFinal}
\end{equation}
and we adopt this strategy also in the present work. A graphical depiction of the online LMS scheme is depicted in Fig.~\ref{fig:LMSScheme}. 

\subsection{Molecular Implementations.}
The goal is now to synthesize eq. (\ref{eq:LMSFinal}) using biochemical reactions. However, in its present form eq. (\ref{eq:LMSFinal}) is incompatible with mass-action rate laws because it contains a negative (i.e., degradation) flux that does not depend on the current value of $\hat{\theta}(t)$. To account for this, we choose the adaptation rate to be proportional to $\hat{\theta}(t)$, i.e., $\tilde{\alpha}(t):= \lambda \hat{\theta}(t)$ and thus,
\begin{equation}
	\frac{\mathrm{d}}{\mathrm{d}t}\hat{\theta}(t) = - \lambda \hat{\theta}(t) X(t) S(t) + \lambda \hat{\theta}(t) Y(t) S(t).
	\label{eq:LMSFinal2}
\end{equation}

While (\ref{eq:LMSFinal2}) is now in principle compatible with mass-action kinetics, it involves trimolecular reactions, that are hard or maybe impossible to realize in practice. However, the trimolecular reaction can be composed from two bimolecular reactions with appropriately chosen rate constants. For example, the reaction $$\mathbf{A}+\mathbf{B}+\mathbf{C} \xrightharpoonup{\lambda} \mathbf{D}$$ can be represented by 
\begin{equation}
\begin{split}
	\mathbf{A}+\mathbf{B}\xrightleftharpoons[b]{f} \mathbf{O} \\
 	\mathbf{O} + \mathbf{C} \xrightharpoonup{\lambda b / f} \mathbf{D},
 \end{split}
 \nonumber
\end{equation}
assuming $b, f >> \lambda$.

Specific implementations of the derived LMS adaptation scheme will be given in the subsequent section.


 \section{Case Studies}
 \label{sec:CaseStudies}

In this section we provide several numerical and analytical examples to demonstrate our molecular LMS estimation framework. In all of the examples, we will consider a target process 
\begin{equation}
	\emptyset \xrightharpoonup{\rho} \mathbf{Y} \xrightharpoonup{\phi} \emptyset,
\end{equation}
with $\rho$ and $\phi$ as the process parameters that we aim to identify using an LMS scheme. As indicated earlier, we restrict ourselves to the case of a single unknown parameter, meaning that we either have $\theta=\left\{\rho\right\}$ or $\theta=\left\{\phi\right\}$. 

\subsection{Self-adjusting birth-rate.}
       
We first consider the case where the birth-rate $\rho$ is unknown to the circuit. The goal is to construct a corresponding adaptive circuit $\mathbf{X}$
\begin{equation}
	\emptyset \xrightharpoonup{\hat{\theta}} \mathbf{X} \xrightharpoonup{\phi} \emptyset,
\end{equation}
whose birth-rate $\hat{\theta}$ adapts to that of $\mathbf{Y}$.
To accomplish this, we require a molecular implementation of relation (\ref{eq:LMSFinal}). The first step is to derive the particular form of the sensitivity function $S(t)$ from the dynamics of $\mathbf{X}$. The latter is given by the rate equation (\ref{eq:RRE}) which in this case reads
\begin{equation}
	\frac{\mathrm{d}}{\mathrm{d}t} X(t) = \hat{\theta} - \phi X(t). \label{eq:XOde}
\end{equation}
Differentiating both sides of eq. (\ref{eq:XOde}) with respect to $\hat{\theta}$ yields
\begin{equation}
	\frac{\mathrm{d}}{\mathrm{d}t} \frac{\partial}{\partial \hat{\theta}}X(t) = \frac{\mathrm{d}}{\mathrm{d}t} S(t) =  1 - \phi S(t). \label{eq:SOde}
\end{equation}
Fortunately, this equation is already in the form of a valid rate equation. In particular, it describes the time-evolution of a birth-death process
\begin{equation}
	\emptyset \xrightharpoonup{1} \mathbf{S} \xrightharpoonup{\phi} \emptyset.
\end{equation}
In conjunction with (\ref{eq:LMSFinal}), the overall adaptive system can be implemented through reactions
\begin{equation}
	\begin{split}
		\emptyset &\xrightharpoonup{\theta} \mathbf{Y} \\
		 \mathbf{Y} &\xrightharpoonup{\phi} \emptyset \\
		\hat{\boldsymbol{\theta}} &\xrightharpoonup{1} \hat{\boldsymbol{\theta}} + \mathbf{X} \\
		\mathbf{X} &\xrightharpoonup{\phi} \emptyset \\
		\emptyset &\xrightharpoonup{1} \mathbf{S} \\
		\mathbf{S} &\xrightharpoonup{\phi} \emptyset \\
		\mathbf{S} + \hat{\boldsymbol{\theta}} + \mathbf{X} &\xrightharpoonup{\lambda} \mathbf{S}+ \mathbf{X} \\
		\mathbf{S} + \hat{\boldsymbol{\theta}} + \mathbf{Y} &\xrightharpoonup{\lambda} \mathbf{S} + 2\hat{\boldsymbol{\theta}} + \mathbf{Y}.
	\end{split}
	\label{eq:LMSBirth}
\end{equation}

\begin{figure}[ht!]
\begin{center}
	\includegraphics[width=0.95\columnwidth]{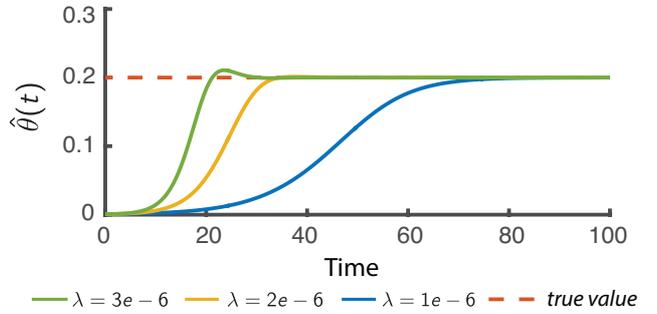}
\caption{Convergence of estimated birth-rate as a function of $\lambda$. The adaptive circuit from (\ref{eq:LMSBirth}) was simulated using parameters $\theta=0.2$ and $\phi=0.01$. Small $\lambda$ lead to slow convergence, while $\lambda>\phi^3/(4 \theta) = 1.25e-6$ cause overshooting and oscillatory behavior.}
\label{fig:AdaptationDet_Lambda}
\end{center}
\end{figure}

We first studied the adaptation performance of (\ref{eq:LMSBirth}) as a function of the tuning parameter $\lambda$ in an idealized noise-free scenario. In this case, the network from (\ref{eq:LMSBirth}) is described by the rate equations
\begin{align}
	\frac{\mathrm{d}}{\mathrm{d}t} Y(t) &= \theta - \phi Y(t)\\
	\frac{\mathrm{d}}{\mathrm{d}t} X(t) &= \hat{\theta}(t) - \phi X(t) \\
	\frac{\mathrm{d}}{\mathrm{d}t} \hat{\theta}(t) &= -\lambda \hat{\theta}(t) X(t) S(t) + \lambda \hat{\theta}(t) Y(t) S(t) \\
	\frac{\mathrm{d}}{\mathrm{d}t} S(t) &=  1 - \phi S(t).
	\label{eq:ODEBirth}
\end{align}
This equation was simulated for different values of $\lambda$ as depicted in Fig.~\ref{fig:AdaptationDet_Lambda}. The results indicate a tradeoff that is associated with the choice of $\lambda$: too small $\lambda$ lead to slow convergence of the scheme, while too large $\lambda$ cause the adaptation scheme to ``overshoot'' the target value and exhibit oscillations. 

In order to analyze the convergence properties of (\ref{eq:LMSBirth}), we performed a local stability analysis of (\ref{eq:ODEBirth}) based on linearization. Noting that $Y(t)$ and $S(t)$ evolve autonomously, we can replace those variables by their steady state values $Y_{\infty}= \theta / \phi$ and $S_{\infty}=1/\phi$, respectively. This yields the reduced system
\begin{align}
\frac{\mathrm{d}}{\mathrm{d}t} X(t) &= \hat{\theta}(t) - \phi X(t) \\
\frac{\mathrm{d}}{\mathrm{d}t} \hat{\theta}(t) &= -\frac{\lambda}{\phi} \hat{\theta}(t) X(t) + \frac{\lambda\theta}{\phi^{2}}\hat{\theta}(t)
\label{eq:ODEBirthReduced}
\end{align}
which has equilibrium points $X_{\infty}^{1}=\hat{\theta}_{\infty}^{1}=0$ and $X_{\infty}^{2}=\theta/ \phi$ and $\hat{\theta}_{\infty}^{2}=\theta$. The respective Jaccobians are given by
\begin{equation}
	A^{1} = \begin{pmatrix}
			-\phi & 1 \\
			 0 & \frac{\lambda \theta}{\phi^{2}}
		\end{pmatrix}
	\quad A^{2} = \begin{pmatrix}
			-\phi & 1 \\
			 -\frac{\theta\lambda}{\phi} & 0
		\end{pmatrix}.
\end{equation}
The eigenvalue $\lambda \theta/\phi^{2}$ of $A^{1}$ is always positive meaning that for initial conditions $\hat{\theta}(0)>0$, the system will not converge to that equilibrium point. From $A^{2}$ we find eigenvalues
\begin{equation}
	\mu_{1} = -\frac{\sqrt{\phi ^4-4 \theta  \lambda  \phi }+\phi ^2}{2 \phi } \quad \mu_{2}=-\frac{-\sqrt{\phi ^4-4 \theta  \lambda  \phi }+\phi ^2}{2 \phi }.
\end{equation}
For any $\lambda>0$, the real part of both eigenvalues is negative, meaning that $\theta$ is the only stable equilibrium of the adaptation scheme. However, we find that for $\lambda>\phi^3/(4 \theta)$, the system will have complex eigenvalues, indicating oscillatory behavior. This can also be seen from in Fig.~\ref{fig:AdaptationDet_Lambda}, especially for the case $\lambda=3e-6$.

\begin{figure}[ht!]
\begin{center}
	\includegraphics[width=0.95\columnwidth]{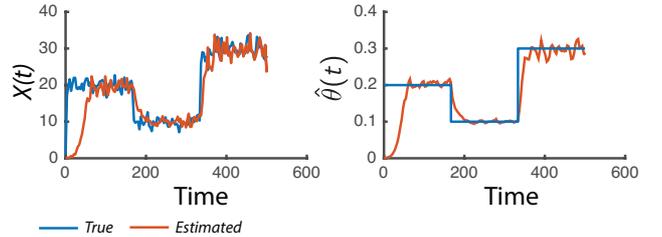}
\caption{Algorithm performance in the presence of molecular noise. The adaptive circuit from (\ref{eq:LMSBirth}) was simulated using the stochastic simulation algorithm to account for molecular noise. We assume that the target value $\theta$ changes spontaneously at certain time points to check the circuit's ability to readapt. The parameters used for the simulations were $\theta=0.2$, $\phi=0.01$ and $\lambda=1e-6$. }
\label{fig:AdaptationNoisy_Jump}
\end{center}
\end{figure}

We next analyzed the circuit's performance in the presence of molecular fluctuations and spontaneous changes in $\theta$. We used the model from eq. (\ref{eq:RTC}) and stochastic simulations \cite{Gillespie2007} to simulate the reaction network from (\ref{eq:LMSBirth}). The results from Fig.~\ref{fig:AdaptationNoisy_Jump} show that the birth-rate $\theta$ and in turn the system output $Y(t)$ is accurately tracked by the molecular LMS scheme. A detailed and quantitative error analysis in the presence of molecular fluctuations will be subject of future work.

\textit{Remark:} We want to point out another interesting property of this particular LMS estimator. Replacing the sensitivity $S(t)$ by a positive constant (e.g., its stationary value $S_{\infty}=1/\phi$) does not change the asymptotic behavior of $\hat{\theta}(t)$, meaning that $\theta$ will remain as the only stable equilibrium point. However, the adaptation law simplifies to
\begin{equation}
\begin{split}
	\frac{\mathrm{d}}{\mathrm{d}t} \hat{\theta}(t) &= -\frac{\lambda}{\phi} \hat{\theta}(t) X(t) + \frac{\lambda}{\phi}\hat{\theta}(t) Y(t) \\
	&= \frac{\lambda}{\phi} \hat{\theta}(t) (Y(t) - X(t)),
\end{split}
\end{equation}
which is structurally equivalent to a specific control motif that has been studied previously \cite{Shoval2010, Briat2016}. In particular, it was shown to act as an integral control circuit exhibiting robust perfect adaptation \cite{Doyle2000}. This points out the potential use of our adaptive estimation framework for studying robustness in biological networks.

\subsection{Self-adjusting death-rate.} We further show how the LMS adaption can be used to identify the target circuit's death rate $\theta=\left\{\phi\right\}$. The corresponding sensitivity $S(t)$ can be shown to satisfy
\begin{equation}
	\frac{\mathrm{d}}{\mathrm{d}t}S(t) = -X(t) - \hat{\theta} S(t).
\end{equation}
There are two issues associated with the above equation. First, it depends on the tuning parameter $\hat{\theta}$, which will change over time due to the adaptation scheme. Correspondingly, the value of $S(t)$ will be different from the actual sensitivity $\frac{\partial}{\partial \hat{\theta}} X(t)$. While this will have an impact on the convergence rate of the LMS scheme, it does not affect its steady state behavior (see analytical results below). The second problem is that $S(t)$ is incompatible with mass-action rate laws due to the negative dependency on $X(t)$. In order to address this problem, we consider the equation for $S^{-}(t)=-S(t)$, which is given by
\begin{equation}
	\frac{\mathrm{d}}{\mathrm{d}t}S^{-}(t) = X(t) - \hat{\theta} S^{-}(t)
\end{equation}
and correspondingly use the LMS update rule
\begin{equation}
\begin{split}
	\frac{\mathrm{d}}{\mathrm{d}t}\hat{\theta}(t) &= - \lambda \hat{\theta}(t) X(t) S(t) + \lambda \hat{\theta}(t) Y(t) S(t) \\
	&= \lambda \hat{\theta}(t) X(t) S^{-}(t) - \lambda \hat{\theta}(t) Y(t) S^{-}(t).
\end{split}
\end{equation}

Overall, the adaptive circuit is given by the reactions
\begin{equation}
	\begin{split}
		\emptyset &\xrightharpoonup{\rho} \mathbf{Y} \\
		 \mathbf{Y} &\xrightharpoonup{\theta} \emptyset \\
		\emptyset &\xrightharpoonup{\rho} \mathbf{X} \\
		\hat{\boldsymbol{\theta}} + \mathbf{X} &\xrightharpoonup{1} \hat{\boldsymbol{\theta}} \\
		\mathbf{X} &\xrightharpoonup{1} \mathbf{X} + \mathbf{S} \\
		\hat{\boldsymbol{\theta}} + \mathbf{S} &\xrightharpoonup{1} \emptyset \\
		\mathbf{S} + \hat{\boldsymbol{\theta}} + \mathbf{Y} &\xrightharpoonup{\lambda} \mathbf{S} + \mathbf{Y} \\
		\mathbf{S} + \hat{\boldsymbol{\theta}} + \mathbf{X} &\xrightharpoonup{\lambda} \mathbf{S}+ 2\hat{\boldsymbol{\theta}} + \mathbf{X}.
	\end{split}
	\label{eq:LMSDeath}
\end{equation}
 
Similar to the previous section, we performed simulations to check the adaptation performance of the circuit as a function of $\lambda$ under idealized noise-free conditions. This allows us to describe the adaptive circuit by the differential equations
\begin{align}	
	\frac{\mathrm{d}}{\mathrm{d}t} Y(t) &= \rho - \theta Y(t)\\
	\frac{\mathrm{d}}{\mathrm{d}t} X(t) &= \rho - \hat{\theta}(t) X(t) \\
	\frac{\mathrm{d}}{\mathrm{d}t} \hat{\theta}(t) &= -\lambda \hat{\theta}(t) X(t) S(t) + \lambda \hat{\theta}(t) Y(t) S(t) \\
	\frac{\mathrm{d}}{\mathrm{d}t} S(t) &=  -X(t) - \hat{\theta}(t) S(t).
	\label{eq:ODEDeath}
\end{align}
 
\begin{figure}[ht!]
\begin{center}
	\includegraphics[width=0.96\columnwidth]{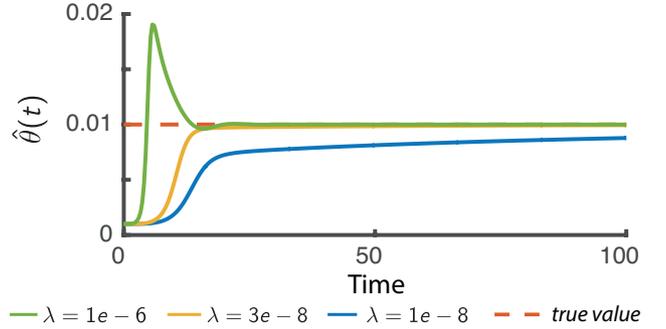}
\caption{Convergence of estimated death-rate as a function of $\lambda$. The adaptive circuit from (\ref{eq:LMSDeath}) was simulated for $\theta=0.1$ and $\phi=0.01$ and different values of $\lambda$. For $\lambda> \theta ^4/ (4 \rho ^2)=2.5e-7$, the adaptation scheme exhibits oscillatory behavior.}
\label{fig:AdaptationDeath}
\end{center}
\end{figure}

We again performed a local stability analysis of the differential equations to investigate the convergence of the circuit. In this case, the sensitivity $S(t)$ is coupled to $X(t)$ and $\hat{\theta}(t)$ and thus, has to be included in the dynamic analysis. After eliminating the equation corresponding to $Y(t)$, we obtain a three-dimensional system
\begin{align}	
	\frac{\mathrm{d}}{\mathrm{d}t} X(t) &= \rho - \hat{\theta}(t) X(t) \\
	\frac{\mathrm{d}}{\mathrm{d}t} \hat{\theta}(t) &= -\lambda \hat{\theta}(t) X(t) S(t) + \frac{\lambda\rho}{\theta} \hat{\theta}(t)  S(t) \\
	\frac{\mathrm{d}}{\mathrm{d}t} S(t) &=  -X(t) - \hat{\theta}(t) S(t),
	\label{eq:ODEDeathReduced}
\end{align}
which has equilibria at the origin and at the point $X_{\infty}^{3} = \rho / \theta  , \hat{\theta}_{\infty}^{3}= \theta$ and $S_{\infty}^{3}= -1/\theta^{2}$. For compactness, we skip explicit expressions of the respective Jaccobian matrices and eigenvalues. However, as in the previous example, we found that for any $\lambda>0$, only the non-zero equilibrium point is stable. For any $\lambda> \theta ^4/ (4 \rho ^2)$, the adaptation exhibit oscillatory behavior, which should be taken into consideration when designing this circuit. These results are confirmed in Fig.~\ref{fig:AdaptationDeath}, for which we simulated (\ref{eq:ODEDeath}) for three different values of $\lambda$.

 \addtolength{\textheight}{-11.80cm}   

\section{CONCLUSIONS}
\label{sec:Conclusions}

LMS schemes provide a powerful and versatile framework for adaptive estimation. In this work we have shown how simple LMS-type estimators that usually run on a computer can be implemented biochemically for the purpose of synthetic biology. Such algorithms would allow a circuit to make inference about its environment and facilitate adaptive behavior. We have shown by simulation that the LMS circuit is able to accurately estimate and track unknown parameters of a birth-death process. We are currently extending the proposed scheme to more general scenarios when multiple parameters are to be estimated simultaneously and when both $\mathbf{X}$ and $\mathbf{Y}$ are arbitrary, possibly multivariate stochastic circuits.  

We anticipate several important potential applications of the presented framework. In \cite{Zechner2016}, optimal filters (such as the Kalman filter \cite{Kalman1961}) were employed to design noise-cancelling synthetic circuits. To this end, the LMS approach provides an attractive alternative to optimal filtering due to its generic and simple structure. In the future, we will extend the approach to nonlinear and multivariate models and provide an in-depth analysis of its properties. 

 
\subsection*{ACKNOWLEDGMENTS.}
This project was financed with a grant from the Swiss SystemsX.ch initiative, evaluated by the Swiss National Science Foundation.

\subsection*{COPYRIGHT INFORMATION.}
This article was presented and published at the 2016 IEEE 55th Conference on Decision and Control (CDC) in Las Vegas. In the original version, Figure 1 was unreferenced and $\theta$ in eq. (28) should have been $\hat{\theta}$. In the present version, both issues have been corrected.




\end{document}